# Valley-polarized domain wall magnons in 2D ferromagnetic bilayers

Doried Ghader

College of Engineering and Technology, American University of the Middle East, Eqaila, Kuwait

doried.ghader@aum.edu.kw

**Abstract**

Valleytronics is a pioneering technological field relying on the valley degree of freedom to achieve novel electronic functionalities. Topological valley-polarized electrons confined to domain walls in bilayer graphene were extensively studied in view of their potentials in valleytronics. Here, we study the magnonic version of domain wall excitations in 2D honeycomb ferromagnetic bilayers (FBL) with collinear order. In particular, we explore the implications of Dzyaloshinskii-Moriya interaction (DMI) and electrostatic doping (ED) on the existence and characteristics of 1D magnons confined to layer stacking domain walls in FBL. The coexistence of DMI and ED is found to enrich the topology in FBL, yet the corresponding domain wall magnons do not carry a well-defined valley index. On the other hand, we show that layer stacking domain walls in DMI-free FBL constitute 1D channels for ballistic transport of topological valley-polarized magnons. Our theoretical results raise hope towards magnon valleytronic devices based on atomically thin topological magnetic materials.

**Introduction**

The recent realization of two-dimensional (2D) materials with intrinsic magnetism [1, 2] constituted an important breakthrough in condensed matter physics. This discovery immediately stimulated tremendous interest in the fundamental physics underlying 2D magnets, as well as their technological potentials [3-22]. With the novel characteristics of their magnetic excitations, 2D magnets are expected to open new horizons in technological fields like magnonics.

The magnon spectrum in monolayer honeycomb ferromagnets mimics the electronic band structure of graphene [11], with degenerate bands at the non-equivalent valleys $\pm K$ (Fig. 1a). The energy degeneracy at the valleys is protected by inversion and time reversal symmetries. Next-nearest-neighbor Dzyaloshinskii-Moriya interaction (DMI), present in several 2D ferromagnets, breaks the time-reversal symmetry [5] and opens topological gaps at $\pm K$ (Fig. 1b). Similar to monolayers, bilayer ferromagnets exhibit a pair of degenerate bands at the inequivalent valleys. Valley gaps can be induced in FBL by intralayer DMI [6, 9] (Fig. 1c) or by breaking the inversion symmetry through layer-dependent electrostatic doping [23, 24] (Fig. 1d). In a recent theoretical study [24], AB-stacked FBL gapped by electrostatic doping (ED) is predicted to be a topological insulating phase, featuring magnon valley currents and Hall effect. This novel behavior is guaranteed by the non-trivial Berry curvature and the no-valley mixing symmetry.

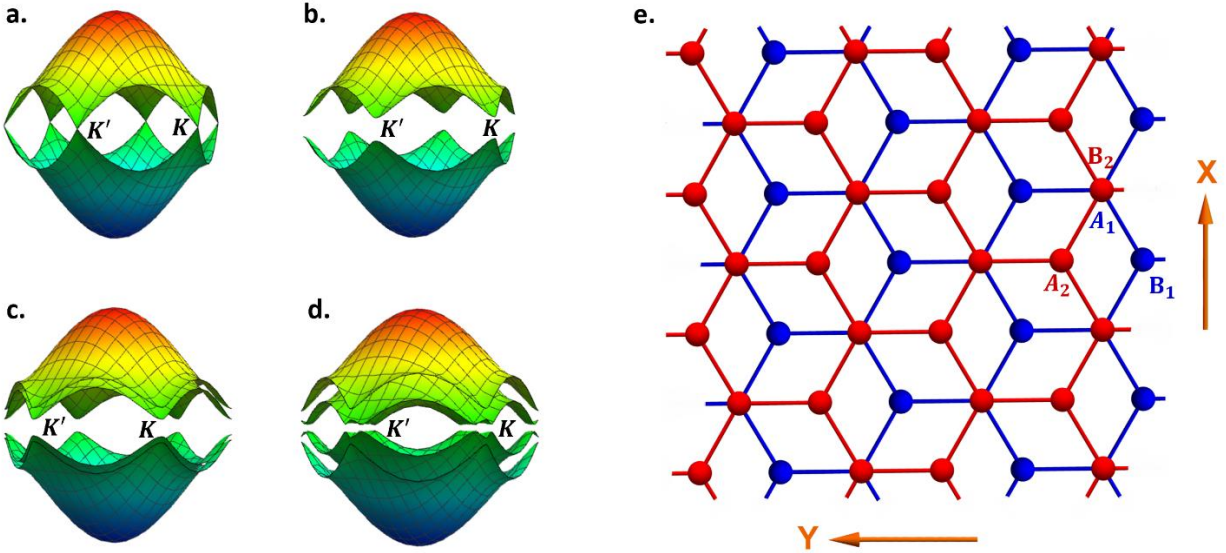

**Figure 1:** (a) and (b) Gapless magnon spectrum and DMI induce gap in monolayer honeycomb ferromagnet. (c) and (d) illustrate the gapped FBL magnon spectrum due to intralayer DMI and layer-dependent electrostatic doping respectively. (e) Schematic representation illustrating the geometry of an AB-stacked FBL.

The valley degree of freedom in electronic systems is proved very promising for quantum technologies, leading to the novel technological field of valleytronics [25–38]. In particular, topological confinement of valley-polarized electronic states on domain walls of bilayer graphene (BLG) has been successfully proposed to realize novel nano-electronic devices. Applying a perpendicular electric field in BLG breaks inversion symmetry and consequently opens a (valley) band gap that admits a topological interpretation [26, 39-46]. The applied field induces a non-trivial Berry curvature for the low energy valence band, with opposite Berry curvature sign near the two valleys. The Berry curvature integral $N_v$ within a single valley $v$ is interpreted as "valley Chern number", converging to quantized values $N_v = \pm 1$ for nonequivalent valleys [26, 27, 47-51]. Novel domain walls can then be formed by a sign reversal of the perpendicular electric field, which flips the sign of $N_v$ and induces a topological invariant $\Delta N_v = \pm 2$ across the electric-field domain wall (EFW) [47]. In agreement with the bulk-edge correspondence, $\Delta N_v$ is associated with the appearance of a pair of valley-polarized chiral electrons confined to the EFW [26, 47]. Topological domain walls for valley transport can also be formed by interlayer stacking reversal in BLG, where the layer stacking switches from AB to BA configurations [27, 33, 52].

The technological interest in the valley degree of freedom was extended to 2D photonics [53] and phononics [54]. Magnon valleytronics, however, did not yet receive the deserved attention. The realization of magnon valleytronic devices demands novel magnetic materials where magnons with preserved valley index can ballistically propagate with lowest dissipation. In this work, we investigate the effects of ED and DMI on the topology and valley-polarization of bulk and LSW magnons in FBL. Perspectives for realizing magnon valleytronic transport are discussed based on the results.

**Methods**

**Four-band formalism.** We start with an AB-stacked honeycomb FBL with nearest neighbor exchange, weak DMI and electrostatic doping. The FBL is composed of 4 sublattices denoted $\{A_1, B_1, A_2, B_2\}$, where subscripts 1 and 2 correspond to layers 1 and 2 respectively. A schematic representation illustrating the geometry and the choice of axes is presented in Fig.1e. The FBL is assumed in a collinear ground state, guaranteed by a sufficiently weak DMI compared to the exchange coupling [5, 6, 9, 17].

The full Brillouin zone (BZ) Hamiltonian for magnons in AB-stacked FBL can be written as [6, 24]

$$\mathcal{H}_{AB} = JS \begin{pmatrix} (3+U)I + \mathcal{H}_1 & \mathcal{H}_3 \\ \mathcal{H}_3^\dagger & (3-U)I + \mathcal{H}_2 \end{pmatrix} \quad (1)$$

with

$$\mathcal{H}_1 = \begin{pmatrix} f_D(\vec{p}) + v_0 & -f(\vec{p}) \\ -f^*(\vec{p}) & -f_D(\vec{p}) \end{pmatrix}, \quad \mathcal{H}_2 = \begin{pmatrix} f_D(\vec{p}) & -f(\vec{p}) \\ -f^*(\vec{p}) & -f_D(\vec{p}) + v_0 \end{pmatrix}, \quad \mathcal{H}_3 = \begin{pmatrix} 0 & -v_0 \\ 0 & 0 \end{pmatrix} \text{ and } I = \begin{pmatrix} 1 & 0 \\ 0 & 1 \end{pmatrix}.$$

The exchange and DMI lattice functions are $f(\vec{p}) = e^{\frac{ip_y}{\sqrt{3}}} + 2\,e^{-\frac{ip_y}{2\sqrt{3}}} \cos(p_x/2)$ and $f_D(\vec{p}) = (D/J)[4\cos(\sqrt{3}p_y/2)\sin(p_x/2) - 2\sin(p_x)]$ respectively. The parameter $v_0 = J_\perp/J$, where $J$ and $J_\perp$ respectively denote the nearest neighbor intra and interlayer exchange coefficients. $S$ denotes spin, $D$ is the DMI coefficient, and $U$ represents the layer-dependent electrostatic doping potential (normalized by $JS$). $p_x$ and $p_y$ are the momenta along $x$ and $y$. The BA-configuration Hamiltonian $\mathcal{H}_{BA}$ can be deduced from $\mathcal{H}_{AB}$ in a straightforward manner.

Worth noting that the magnonic Hamiltonian $\mathcal{H}_{AB}$ (equivalently $\mathcal{H}_{BA}$) does not have an identical electronic analogue and is candidate for new physics. Even in the absence of DMI, the magnonic Hamiltonian differs from that of bilayer graphene, particularly due to the presence of $v_0$ on the diagonal terms corresponding to $A_1$ and $B_2$ dimer sites (see diagonal terms in $\mathcal{H}_1$ and $\mathcal{H}_2$). These onsite potentials has observable effects on the magnon spectrum and the corresponding eigenfunctions.

We next expand $\mathcal{H}_{AB}$ near $\pm K$ valleys which yields the $4 \times 4$ Dirac Hamiltonians

$$\mathcal{H}_{AB}^{K} = JS\left[\tau_0\left(3\sigma_0 + m\sigma_z + \tfrac{\sqrt{3}}{2}\vec{p}\cdot\vec{\sigma}\right) + U\tau_z\sigma_0 + v_0\Gamma_{AB}\right] \tag{2a}$$

$$\mathcal{H}_{AB}^{-K} = JS\left[\tau_0\left(3\sigma_0 + m\sigma_z - \tfrac{\sqrt{3}}{2}\vec{p}\cdot\vec{\sigma}^*\right) + U\tau_z\sigma_0 + v_0\Gamma_{AB}\right] \tag{2b}$$

with $m = 3\sqrt{3}\,D/J$. The $2 \times 2$ Pauli matrices, $\sigma_i$ and $\tau_i$, are defined in the sublattice and layer spaces respectively. Tensor products are implemented between $\sigma$ and $\tau$. The matrix $\Gamma_{AB} = \tfrac{1}{2}(\tau_z\sigma_z + \tau_0\sigma_0) - \tfrac{1}{4}(\tau_+\sigma_+ + \tau_-\sigma_-)$, to be replaced by $\Gamma_{BA} = \tfrac{1}{2}(-\tau_z\sigma_z + \tau_0\sigma_0) - \tfrac{1}{4}(\tau_+\sigma_- + \tau_-\sigma_+)$ for BA-stacking. Note that the matrices $\tfrac{1}{2}(\pm\tau_z\sigma_z + \tau_0\sigma_0)$ in $\Gamma_{AB}$ and $\Gamma_{BA}$ are consequences of the interlayer exchange onsite potentials discussed above.

We next consider a 2D FBL with LSW (at $x = 0$), separating two regions with AB and BA stacking in the domains $x > 0$ and $x < 0$ respectively. The corresponding four-components magnon wavefunctions in the regions $x \gtrless 0$ are denoted $\psi^\pm$, with $\psi^\pm = \{\phi_i^\pm, i = 1, \ldots, 4\}$. The translation symmetry in the system is broken (preserved) along the $x-$direction ($y-$direction). The behavior of $\psi^\pm$ is hence dictated by the functional form $e^{-\lambda x}e^{ip_y}$, where $\lambda = \alpha + i\beta$ denotes the generalized phase factor in the $x-$direction [15, 18, 20, 21, 26, 27]. For intragap modes confined to the LSW, $\lambda$ is real and the wavefunction decays exponentially away from the LSW.

Reinstating the momenta as differential operators in $\mathcal{H}_{AB}^K$ (equivalently $\mathcal{H}_{BA}^K$), the wave equation $(\mathcal{H}_{AB}^K - \epsilon I)\psi = 0$ yields four real solutions for $\lambda$, namely

$$\lambda_{1(2)}^\pm = \pm\sqrt{p_y^2 - \frac{a+(-)b}{v^2}} \tag{3}$$

with $a = 9 - m^2 + \Omega^2 + 3v_0 - \Omega(6 + v_0) + U^2$, $\Omega = \frac{\epsilon}{JS}$, and $b = \{(-3+\Omega)[(-3+\Omega)v_0^2 + 4(-3+\Omega)U^2 - 4v_0U(m+U)]\}^{1/2}$.

The solutions $\lambda^\pm$ are adopted for $\psi^\pm$ respectively, which yields the functional forms $\phi_i^\pm = u_{i1}e^{-\lambda_1^\pm x}e^{ip_y} + u_{i2}e^{-\lambda_2^\pm x}e^{ip_y}$. The four-band formalism hence generates 16 coefficients $\{u_{ij}; i = 1,..4, j = 1,2\}$, to be determined via boundary conditions derived using the matching method [26, 27, 55-59].

**Boundary conditions.** The boundary conditions are derived by matching the wavefunctions components $\phi_i^\pm$ and their derivatives across the LSW. First, the magnon wavefunction is required to be continuous across the domain wall, resulting in 4 boundary equations $\phi_i^+(x = 0^+) = \phi_i^-(x = 0^-)$. The remaining 12 boundary equations are derived by matching the rows of the wave equations $(\mathcal{H}_{AB}^K - \epsilon I)\psi^> = 0$ and $(\mathcal{H}_{AB}^K - \epsilon I)\psi^< = 0$ across the LSW. For example, considering the first row from each equation yields

$$v_0 \phi_1^+ - i\frac{\sqrt{3}}{2}(\partial_x \phi_2^+ - \partial_x \phi_2^-) - v_0 \phi_4^+ = 0 \tag{4a}$$

$$(3 + U + m - \Omega)(\partial_x \phi_1^+ - \partial_x \phi_1^-) + v_0(\partial_x \phi_1^+ - \partial_x \phi_4^+) - i\frac{\sqrt{3}}{2} p_y(\partial_x \phi_2^+ - \partial_x \phi_2^-)$$
$$- i\frac{\sqrt{3}}{2}(\partial_x^2 \phi_2^+ - \partial_x^2 \phi_2^-) = 0 \tag{4b}$$

$$(3 + U + m - \Omega)(\partial_x^2 \phi_1^+ - \partial_x^2 \phi_1^-) + v_0(\partial_x^2 \phi_1^+ - \partial_x^2 \phi_4^+) - i\frac{\sqrt{3}}{2} p_y(\partial_x^2 \phi_2^+ - \partial_x^2 \phi_2^-)$$
$$- i\frac{\sqrt{3}}{2}(\partial_x^3 \phi_2^+ - \partial_x^3 \phi_2^-) = 0 \tag{4c}$$

The first equation is determined by subtracting the rows. Applying $\partial_x$ and $\partial_x^2$ followed by subtraction yields the second and third equations respectively.

With the boundary conditions in hand, we substitute $\phi_i^\pm$ and deduce a $16 \times 16$ linear system, $M|u\rangle = 0$. The determinant equation $|M(p_y, \Omega)| = 0$ then yields the dispersion relations $\Omega(p_y)$ for the intragap magnon modes confined to the LSW.

We succeeded in reducing the determinant equation, which boils down to

$$(c_1 \lambda_1^+ + c_2 \lambda_2^+ + c_{12} \lambda_1^+ \lambda_2^+ + c_0)(c_1 \lambda_1^+ + c_2 \lambda_2^+ - c_{12} \lambda_1^+ \lambda_2^+ - c_0) = 0 \tag{5}$$

with

$c_{1(2)} = \frac{\sqrt{3}}{2} v_0 \left[-9 + 3a + m^2 - U^2 - \frac{3}{2} p_y^2 - 6v_0 - (+)b - (+)(6 - e + 2v_0)\Omega\right]$ and $c_{12} = \frac{3}{2}\left(-a + \frac{3}{4} p_y^2\right)$.

**Results**

The full BZ Hamiltonians derived in the previous section are particularly useful to calculate the bulk magnon bands, their Berry curvatures and Chern numbers. They further give access to the magnon thermal Hall and Nernst conductivities induced by topological bands in FBL. We will denote the 4 energy bands of $\mathcal{H}_{AB}(\vec{p})$ by $[\epsilon_4, \epsilon_3, \epsilon_2, \epsilon_1]$, arranged in descending energy order. These can be equivalently derived using $\mathcal{H}_{BA}(\vec{p})$.

Without ED, the DMI induces a non-trivial Berry curvature with 6-fold rotational symmetry for the four bands [6, 9]. The corresponding band Chern numbers are $[0, -2, 0, 2]$, which are independent of the exact value of DMI and interlayer exchange. Due to the 6-fold rotational symmetry, the $\pm K$ valleys are topologically identical. Moreover, the magnon Berry curvatures are independent of the stacking, and the AB-BA domain walls do not confine any magnon modes.

The band topology becomes more exotic in the full $(v_0, U, D)$ parametric space of $\mathcal{H}_{AB}(\vec{p})$. We have numerically investigated the magnon band topology all over the parametric space region $\mathcal{R} = \{(v_0, U, D), \ 0.1 \leq v_0 \leq 0.5, 0.02J \leq D \leq 0.2J, 0.05 \leq U \leq 0.3\}$. This region is suitable for the general assumption of FBL with weak DMI and interlayer exchange. Our numerical investigation demonstrates that varying one of the Hamiltonian's parameters at a time (interlayer exchange, DMI or ED) induces topological phase transitions. Consequently, we were able to identify four different topological phases with Chern numbers $[0, -2, 1, 1]$, $[-1, -1, 1, 1]$, $[0, 0, -1, 1]$, and $[0, -2, 0, 2]$. In other words, region $\mathcal{R}$ is divided into four subregions corresponding to these topological phases. As a sample of our numerical results, Figs, 2a-2d present the AB-configuration Berry curvatures of valence-like ($\epsilon_2$) band for (randomly) selected values of ED, DMI and interlayer exchange in the four subregions. The Berry curvatures for the rest of the bands (not presented) show a similar profile. The Berry curvatures are calculated from the full BZ Hamiltonian $\mathcal{H}_{AB}$ and using the numerical approach developed in reference [60]. Cases 2a-2c correspond to the three new topological phases, with Chern numbers $[0, -2, 1, 1]$, $[-1, -1, 1, 1]$, and $[0, 0, -1, 1]$ respectively. Case 2d reproduces the topological phase with Chern numbers $[0, -2, 0, 2]$, despite the presence of ED. Further analysis on the observed topological phase transitions will be presented elsewhere.

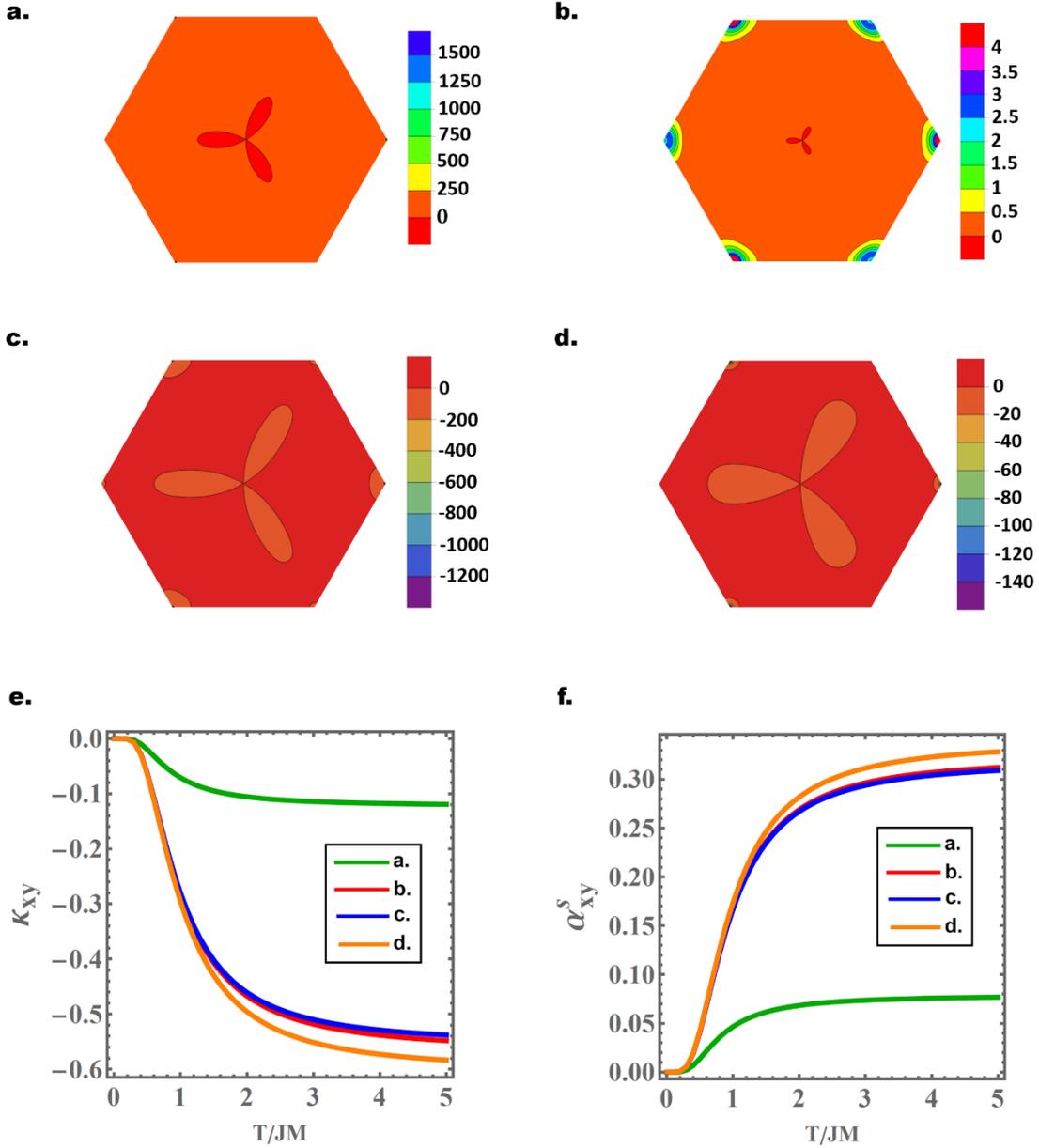

**Figure 2:** (a-d) Sample of the numerically calculated Berry curvatures for $\epsilon_2$ band in FBL with ED and DMI. The parameter choices are $(U = 0.1, D = 0.05J, \ v_0 = 0.3)$, $(U = 0.2, D = 0.05J, \ v_0 = 0.2)$, $(U = 0.2, D = 0.02J, \ v_0 = 0.3)$, and $(U = 0.1, D = 0.05J, \ v_0 = 0.5)$ respectively. The magnon spectra corresponding to these choices are topological, with Chern numbers $[0, -2, 1, 1]$, $[-1, -1, 1, 1]$, $[0, 0, -1, 1]$, and $[0, -2, 0, 2]$ respectively. (e) and (f) present the temperature evolution of Hall ($\kappa_{xy}$) and Nernst ($\alpha^s_{xy}$) conductivities corresponding to the parameters in (a-d).

Figures 2a-2d further illustrate the Berry curvature valley mixing induced by the DMI. This is a general effect for all four bands and is independent of the particular values of the parameters. Consequently, valley Chern numbers and Hall effect [26, 47] cannot be defined in electrostatically

doped FBL with nonzero DMI, and valley-polarized transport of bulk or domain wall magnons cannot be realized. Nevertheless, the non-trivial topology of the bands induces standard Hall and Nernst conductivities [3, 4, 6, 7, 12, 17, 22]. These are plotted in figures 2e and 2f for the four different topological phases with same parameter choice as Figs. 2a-2d. The conductivities show a standard profile, with no thermal excitations at $T = 0$ and saturation at high temperatures. The Hall ($\kappa_{xy}$) and Nernst ($\alpha_{xy}^S$) conductivities are calculated numerically using the standard equations,

$$\kappa_{xy} = -\frac{k_B^2 T}{\hbar V} \sum_{\vec{p},i} c_2\left(g\big(\epsilon_i(\vec{p})\big)\right) \mathcal{B}_i(\vec{p})$$

(6a)

$$\alpha_{xy}^S = \frac{k_B}{V} \sum_{\vec{p},i} c_1\left(g\big(\epsilon_i(\vec{p})\big)\right) \mathcal{B}_i(\vec{p})$$

(6b)

Here, $\mathcal{B}_i$ represents the Berry curvature for band $i$ and momentum $\vec{p}$ is summed over the entire BZ. The function $g(\epsilon_i) = \left[e^{\epsilon_i/k_B T} - 1\right]^{-1}$ is the Bose-Einstein distribution function, $c_1(x) = (1+x)\ln(1+x) - x\ln x$, and $c_2(x) = (1+x)\left[\ln\left(\frac{1+x}{x}\right)\right]^2 - (\ln x)^2 - 2\text{Li}_2(-x)$. Also $\text{Li}_2$ stands for the dilogarithm function, $V$ the volume of the system and $k_B$ the Boltzmann constant. We set $k_B = \hbar = 1$ in our numerical calculation.

Another important consequence of ED is that it breaks inversion symmetry and hence the 6-fold rotational symmetry of the Berry curvatures. This gives relevance to the AB-BA domain walls in FBL with ED and DMI. As illustrated numerically in figure 3a, inverting the layer registry rotates the Berry curvatures by $\pi/3$. The LSW hence separates two regions with different Berry curvatures but identical Chern numbers. Although the LSW is topologically compensated, it can still host 1D magnon modes (figures 3b and 3c). Indeed, the localized modes in the present case do not correspond to a topological invariant valley index, and the LSW spectrum is completely gapped near $\pm K$ valleys. Despite some similarities with topologically compensated LSW in BLG with reversed gating and layer stacking [27], the present case is topologically different and does not have an identical fermionic analogue.

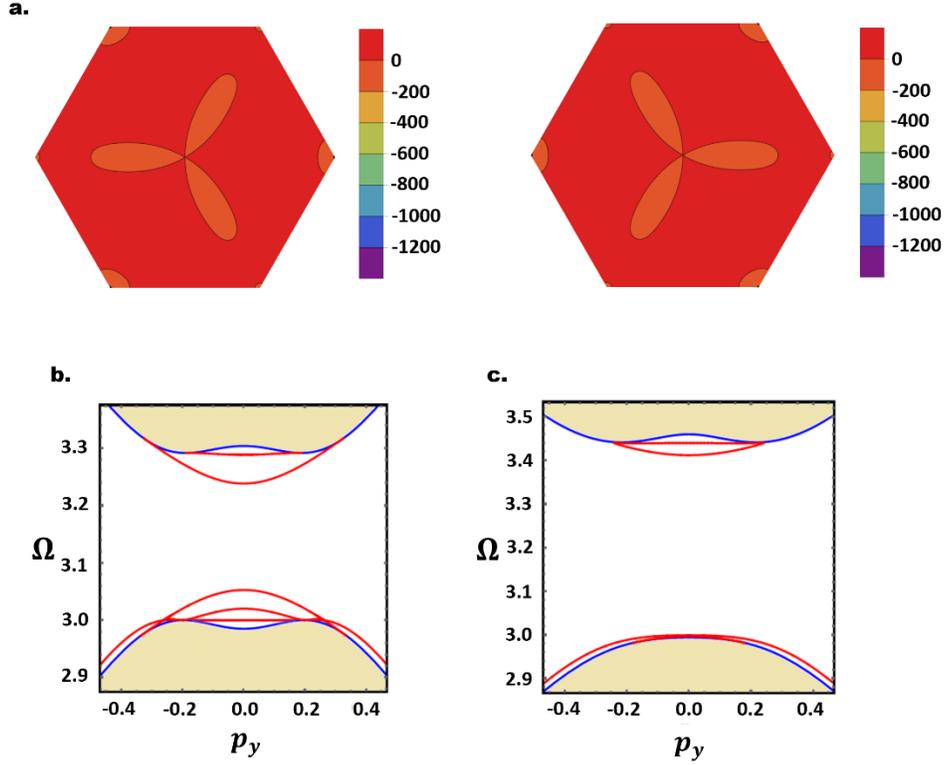

**Figure 3:** (a) Berry curvatures for the $\in_2$ band in AB (left) and BA (right) stacked FBL with parameters $(U = 0.2, D = 0.02J, v_0 = 0.3)$. (b) and (c) show the magnon bulk spectra (shaded regions) and domain wall magnon modes (red) near the $K-$ valley for $(U = 0.2, D = 0.02J, v_0 = 0.3)$ and $(U = 0.2, D = 0.05J, v_0 = 0.3)$ respectively.

The last case to investigate in our search for valley-polarized magnons is FBL with ED and negligible DMI. This system is the closest analogue to bilayer graphene, with the ED in FBL playing the role of the perpendicular electric field in bilayer graphene (both break the inversion symmetry). As stated previously, however, the magnonic Hamiltonian is different due to the onsite interlayer exchange potentials. In addition to their effects on the bulk magnon energy and wavefunctions, these onsite terms contribute significantly to the decay factors (Eq.3) and the energy of domain wall magnons (Eqs.5). Despite these observable effects, numerical investigation on the $D = 0$ projection of $\mathcal{R}$ demonstrates that the magnon Berry curvatures, similar to their fermionic counterparts, are characterized by the no valley mixing symmetry and has opposite sign near the inequivalent valleys. A sample of our numerical results illustrating the Berry curvatures profiles is presented in Fig.4. Moreover, integrating the Berry curvature of the conduction-like band within the $vK$ valley yields a quantized valley Chern number $N_v = \pm 1$, in a remarkable analogy with BLG [26, 27, 47-51]. The valley index hence represents a good quantum number in the present case. This conclusion is general and independent of the numerical values of the parameters $v_0$ and $U$.

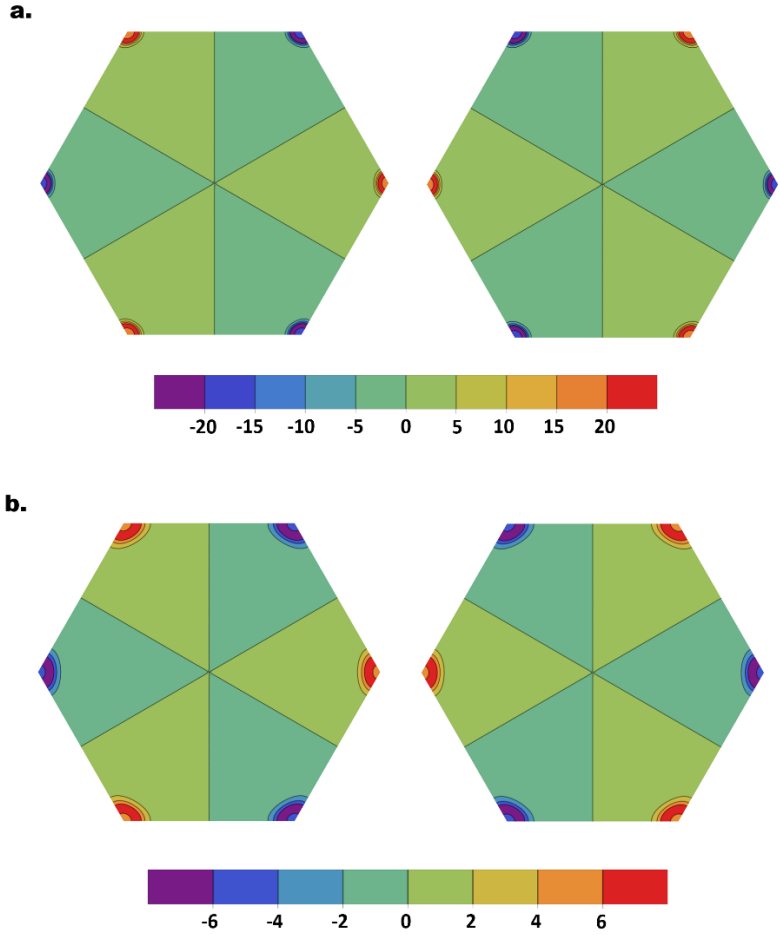

**Figure 4:** (a) and (b) Berry curvatures for the conduction-like band $\epsilon_3$ in AB (left) and BA (right) FBL, with ($U = 0.2$, $v_0 = 0.2, D = 0$) and ($U = 0.3$, $v_0 = 0.5, D = 0$).

We next consider the effect of LSW on the valley Chern numbers. For a uniform ED potential throughout the sample, reversing the stacking configuration results in a $\pi/3$ rotation of the Berry curvature, associated with a sign reversal of $N_v$ (Fig. 4). The LSW thus separates two distinct topological phases, and induces a topological invariant $\Delta N_v = \pm 2$ across the LSW [27, 47]. In accordance with the bulk-edge correspondence, a pair of topologically protected valley-polarized chiral magnon modes are expected to flow along the AB-BA interface, with $\pm K$ valley magnons propagating in opposite directions. This is numerically confirmed in figure 5 for selected values of $U$ and $v_0$. Regardless of the number of intragap modes, there are only two chiral modes per valley, propagating along the domain wall while preserving the valley index. To our knowledge, this constitutes the first theoretical realization of topologically protected 1D magnon valley transport.

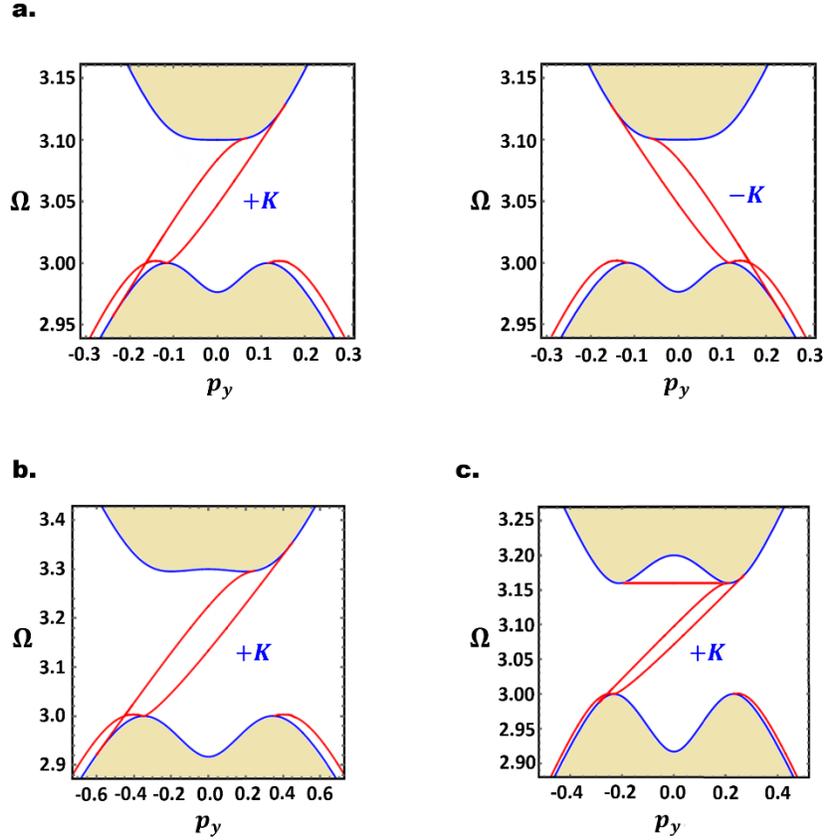

**Figure 5:** (a) Left and right figures respectively show the 1D intragap LSW magnon modes, with 2 chiral modes, near $\pm K -$ valleys. The parameters are $(U = 0.1, \ D = 0, v_0 = 0.2)$. (b) and (c) show only $+K$ valley modes for FBL with $(U = 0.3, D = 0, \ v_0 = 0.5)$ and $(U = 0.2, \ D = 0, \ v_0 = 0.2)$ respectively.

## Discussion

The inequivalent valleys in the BZ of 2D honeycomb ferromagnets provide an additional degree of freedom for magnons, the valley-index, and its manipulation can lead to novel magnonic devices. Topologically protected magnons are robust against various dissipation sources and are candidates to overcome the difficulty in harnessing the magnon valley degree of freedom. Inspired by the intensive research on bilayer graphene for valleytronic applications, we have presented a comprehensive theoretical investigation addressing the effects of ED and DMI on the topology and valley-polarization of bulk and LSW magnons in FBL. Importantly, the presence of DMI in FBL gives them some advantage over bilayer graphene, since the spin-orbit coupling (the electronic analogue of DMI) is absent in the latter. Including the DMI in the formalism, we have predicted several topological phases in electrostatically doped FBL, which are absent in the electronic theory of bilayer graphene. The theory of LSW magnons was also developed beyond the bilayer graphene case, including the DMI and the interlayer exchange onsite potentials. Nonetheless, we have concluded that FBL with both ED and DMI are not useful for valleytronic applications.

Turning off the DMI in FBL results in a magnonic Hamiltonian that can be differentiated from its fermionic version by the interlayer interaction onsite potentials on $A_1$ and $B_2$ dimer sites. Our study proves that LSW in DMI-free electrostatically doped FBL can be used to control the valley degree of freedom. In particular, our study shows that magnon valley Chern numbers are well defined, change sign across LSW, and induce 1D topological boundary magnon states. Similar to electrons in bilayer graphene, the LSW magnon modes have several interesting features that are useful for magnon valleytronics. These modes are gapless and chiral, meaning that their propagation direction along the AB-BA boundary is dictated by the valley-index. The velocity-valley locking and topological protection allows for a long-range ballistic valley transport that conserves the valley-index. Disorder (structural or magnetic) might appear near the domain walls, yet the existence of the valley-polarized 1D boundary modes is topologically guaranteed by the bulk-edge correspondence, as long as the bulk of the sample is not disordered. These predictions open opportunities to realize magnon valleytronic (nano-)devices, including valley valves, filters, waveguides and transistors. Bilayers have already been fabricated from several 2D ferromagnetic materials like $CrBr_3$ [61, 62], $Cr_2Ge_2Te_6$ [1], and others. With the rapid advancement in experimental techniques to fabricate, dope and characterize 2D magnetic materials, experimental testing of our predictions shall be feasible.

**Acknowledgments**